\documentclass[aps,prd,12pt,superscriptaddress]{revtex4-2}
\usepackage{amsmath}
\usepackage{latexsym}
\usepackage{amsfonts}
\usepackage{graphicx}
\usepackage{hyperref}
\usepackage{color}
\usepackage{makecell}
\usepackage{CJK}
\usepackage{comment}

\begin{document}

\title{Constraint on the mass of graviton with gravitational waves}
\author{Qing Gao}
\email{gaoqing1024@swu.edu.cn}
\affiliation{School of Physical Science and Technology, Southwest University, Chongqing 400715, China}

\begin{abstract}
We consider the effects of the mass of graviton on both the waveform of gravitational waves (GWs) and the antenna response to GWs. We determine that the effect on the response function is negligible for small mass. Using the Fisher matrix method, we perform parameter estimations with space-based GW detectors for massive binary black holes (BBHs) in massive gravity theory. The wavelength of massive graviton can be constrained to be $\lambda_g >  1.91\times 10^{19}$ m and the mass can be constrained to be $m_g < 1.16\times 10^{-61}$ kg by 1-year observation of equal-mass massive BBHs with space-based GW detectors.
\end{abstract}

\maketitle

\section{Introduction}

Although Einstein’s theory of general relativity (GR) achieved tremendous success in explaining gravitational and cosmological phenomena, the problems of quantum gravity and spacetime singularity, the existence of dark matter, the acceleration of the universe, and the physics of black holes indicate that we need to modify the theory of gravity. The alternative theories of gravity include the Brans–Dicke  theory \cite{Brans:1961sx}, Lovelock gravity \cite{Lovelock:1971yv}, Horndeski theory \cite{Horndeski:1974wa} and its generalizations \cite{Deffayet:2009mn}, general nonlinear $f(R)$ gravity \cite{Buchdahl:1970ynr}, tensor–vector–scalar theory (or T$e$V$e$S theory) \cite{Bekenstein:2004ne} and generalized T$e$V$e$S theory \cite{Seifert:2007fr}, Einstein-\ae ther theory \cite{Jacobson:2000xp,Jacobson:2004ts}, dynamical Chern–Simons gravity \cite{Jackiw:2003pm}, Ho\v{r}ava gravity \cite{Horava:2009uw}, Dvali, Gabadadze, and Porrati (DGP) gravity \cite{Dvali:2000hr}, bimetric theory of gravity \cite{Rosen:1940zz,Rosen:1973zz}, linearized Fierz–Pauli (FP) massive gravity \cite{Fierz:1939ix} and its generalization \cite{Gambuti:2021meo} without the van Dam, Veltman, and Zakharov (vDVZ) discontinuity \cite{vanDam:1970vg,Zakharov:1970cc}, and de Rham, Gabadadze, and Tolley (dRGT) massive gravity \cite{deRham:2010kj} that is free from the Boulware–Deser ghost \cite{Boulware:1972zf,Hassan:2011hr}. For a review of alternative theories of gravity and massive gravity, please see Refs. \cite{Hinterbichler:2011tt,Clifton:2011jh,deRham:2014zqa}. In this study, we focus on the mass of graviton. The constraint on the mass of graviton from the new solution of the ephemeris INPOP19a  is $m_g\leq 3.16\times 10^{-23}$ eV/$c^2$ ($5.63\times 10^{-59}$ kg) at the 90\% confidence level \cite{Bernus:2020szc}. Based on the observations of the change of the orbital period from PSR B1913+16 \cite{Weisberg:2016jye} and PSR J1738+0333 \cite{Freire:2012mg}, the constraints on the mass of graviton in FP gravity \cite{Fierz:1939ix,Gambuti:2021meo} and DGP gravity \cite{Dvali:2000hr} were discussed in Ref. \cite{Poddar:2021yjd}.

In GR, gravitational waves (GWs) propagate at the speed of light with two transverse polarization states. In alternative theories of gravity, GWs may have up to six polarizations and their propagation speed may differ from the speed of light \cite{Eardley:1974nw,Liang:2017ahj,Hou:2017bqj,Gong:2017bru,Gong:2017kim,Gong:2018cgj,Gong:2018ybk,Gong:2018vbo,Hou:2018djz,Oikonomou:2021kql,Oikonomou:2020sij,Odintsov:2020sqy,Oikonomou:2020oil}. The detection of the polarizations of GWs was discussed in Refs. \cite{Nishizawa:2009bf,Hayama:2012au,Isi:2015cva,Isi:2017equ,Callister:2017ocg,DiPalma:2017qlq,LIGOScientific:2018czr,Takeda:2018uai,Takeda:2019gwk,Liu:2020mab,Zhang:2021fha}. Because of the modification of the dispersion relation similar to that in massive gravity theory, the speed of GW depends on its frequency. For coalescing binaries, the frequency of the gravitational radiation sweeps from low frequency to high frequency with time. Thus, GWs emitted in the early inspiral travel more slowly than those emitted close to the merger, whereas GWs emitted at an earlier time may reach the detector later than GWs emitted at a later time, leading to a distortion of the observed phase of GWs. Therefore, the observation of such dephasing can place a constraint on the mass of graviton.

The Laser Interferometer Gravitational-Wave Observatory (LIGO) Scientific Collaboration and the Virgo Collaboration have already detected tens of GW events, and we are in a new era of gravitational universe \cite{Abbott:2016blz,LIGOScientific:2018mvr,LIGOScientific:2020ibl,LIGOScientific:2021usb,LIGOScientific:2021djp,Lu:2022wuk}. With GWs, it is possible to test GR and probe the nature of gravity in the strong-field and nonlinear regions \cite{TheLIGOScientific:2016src,Abbott:2018lct,LIGOScientific:2019fpa,Abbott:2020jks,Zhang:2021fha,Dai:2021olt,Zhang:2022hbt}. The recent progress in GW physics was discussed in Refs. \cite{Cai:2017cbj,Bian:2021ini}. The observations of the first binary neutron star event GW170817 and its electromagnetic counterpart GRB170817A constrain the propagation speed of GW as $-3\times 10^{-15}<v_{gw}/c-1\le 7\times 10^{-16}$ \cite{LIGOScientific:2017zic}. The Bayesian analysis of GW170817 using the TaylorF2 waveform model yielded $m_g\le 1.305\times 10^{-57}$ kg for the low-spin prior \cite{Shoom:2022cmo}. Using 43 GWTC-3 binary black hole (BBH) events, the LIGO, Virgo, and KAGRA collaborations determined the 90$\%$ confidence level of the mass of graviton to be $m_g\leq 1.27\times 10^{-23}$ eV/$c^2$ ($2.26\times 10^{-59}$ kg) \cite{LIGOScientific:2021sio}. Space-based GW detectors, such as LISA \cite{Danzmann:1997hm,LISA:2017pwj}, TianQin \cite{TianQin:2015yph}, and Taiji \cite{Hu:2017mde}, can detect GWs from inspiral to coalescence and ringdown, which may last for years and GW sources at a larger distance. Thus, more stringent constraints on the mass of graviton can be obtained using space-based GW detectors because the accumulated dephasing is larger in long-duration GWs farther away from the observer. For the 1-year observation of equal-mass massive BBH with the component mass $10^7M_\odot$ inspirals at the luminosity distance $D_L = 3$ Gpc using LISA, the Fisher information matrix (FIM) analysis with GW waveform of up to 1.5 post-Newtonian (PN) order yielded as large as $6.9\times 10^{19}$ m lower bound for the graviton Compton wavelength $\lambda_g = h/(m_g c)$ \cite{Will:1997bb}. With an improved noise curve for LISA, the lower bound was revised as $\lambda_g > 4.8\times10^{19}$ m \cite{Will:2004xi}. Using the 2 PN GW waveform for nonprecessing spinning BBHs at $D_L = 3$ Gpc with spin–orbit coupling, the 1-year observation of equal-mass massive BBHs with the component mass $10^7M_\odot$ obtained using LISA yielded the lower bound $\lambda_g>2.2 \times10^{19}$ m \cite{Berti:2004bd}. The Monte Carlo simulations of $10^4$ BBHs randomly distributed and oriented in the sky with the component mass $(10^6+10^6)M_{\odot}$ obtained using LISA yielded the average lower bound $\lambda_g > 1.33\times10^{19}$ m \cite{Berti:2004bd}. For processing eccentric BBHs, the restricted 2 PN waveforms, including the effects of spin–orbit and spin–spin couplings, spin precession, and eccentricity of the orbit, were employed to constrain the mass of graviton \cite{Yagi:2009zm}. The pattern-averaged analysis of the 1-year observation of the inspirals of BBHs with the component masses $(10^7+10^7)M_{\odot}$, $(10^7+10^6)M_{\odot}$, $(10^6+10^6)M_{\odot}$, and $(10^6+10^5)M_{\odot}$ at $D_L = 3$ Gpc obtained using LISA yielded the constraints $\lambda_g > 1.2\times10^{19}$ m, $\lambda_g>4.1\times10^{18}$ m, $\lambda_g > 3.6\times10^{18}$ m, and $\lambda_g > 1.3\times10^{18}$ m, respectively, whereas the Monte Carlo simulations of $10^4$ BBHs randomly distributed and oriented in the sky with the component mass $(10^7+10^6)M_{\odot}$ at $D_L = 3$ Gpc obtained using LISA yielded $\lambda_g > 3.1\times10^{19}$ m on average \cite{Yagi:2009zm}.
For space-based GW detectors, the arm length of the detector is comparable to or even larger than the wavelength of in-band GWs; thus, it is necessary to consider the frequency dependence of the antenna response. For massive gravitons, the dispersion relation affects the frequency-dependent antenna response. In previous analyses of the pattern-averaged response function \cite{Will:1997bb,Will:2004xi,Berti:2004bd,Yagi:2009zm}, the effect of the mass of graviton on the antenna response was not considered. In this study, we revisited the constraint on the mass of graviton by considering the effect of the mass of graviton on the antenna response. We use GW waveforms with both amplitude and phase corrections of up to 3.5 PN and the FIM method to conduct parameter estimation for nonspinning BBH inspirals in massive gravity theory. In this study, we consider tensor modes only. The paper is organized as follows: In Sec.\ref{sec2}, we review the semi-analytical formulas of the averaged response functions of the tensor, vector, breathing, and longitudinal modes in massive gravity theory. Then, we propose analytical approximations for these averaged response functions. In Sec.\ref{sec3}, we discuss the parameter estimation and the constraint on the mass of graviton obtained using LISA, Taiji, and TianQin with the analytical formulas of the averaged response functions for massive GWs using the FIM method. The conclusion is drawn in Sec. \ref{sec4}.

\section{Semi-analytical formulas}
\label{sec2}

For GWs propagating in the direction $\hat{\Omega}$,
\begin{equation}
\label{hijt}
h_{ij}(t)=\sum_{A} e^A_{ij} h_A(t),
\end{equation}
with the signal registered in the GW detector expressed as follows:
\begin{equation}
\label{gwst}
s(t)=\sum_A F^A h_A(t),
\end{equation}
where $e^A_{ij}$ is the polarization tensor and $A=+,\times,x,y,b,l$ are the plus, cross, vector $x$, vector $y$, breathing, and longitudinal polarizations, respectively. The angular response function $F^A$ for the polarization $A$ is expressed as follows:
\begin{equation}
\label{faeq1}
F^A=\sum_{i,j} D^{ij} e^A_{ij}.
\end{equation}
The detector tensor $D^{ij}$ for an equal-arm space-based interferometric detector with a single round-trip light travel is expressed as follows:
\begin{equation}
D^{ij}=\frac{1}{2}[\hat{u}^i \hat{u}^j T(f,\hat{u}\cdot\hat{\Omega})-\hat{v}^i \hat{v}^j  T(f,\hat{v}\cdot\hat{\Omega})],
\end{equation}
where $\hat{u}$ and $\hat{v}$ are the unit vectors along the arms of the detector. For GWs in massive gravity theory, the propagation speed of GWs $v_{gw}(f)=\sqrt{1-(m_g c^2)^2/(hf)^2}$ is different from the speed of light $c$ and depends on the frequency of GWs. Taking this effect into account, the transfer function $T(f,\hat{n}\cdot\hat{\Omega})$ for a single round-trip in the arm is expressed as follows \cite{Tinto:2010hz,Blaut:2015qaa}:
\begin{equation}
\label{transferfunction}
\begin{split}
T(f,\hat{n}\cdot\hat{\Omega})=\frac{1}{2}& \left\{ \text{sinc}\left[\frac{f}{2f^*}(1-\hat{n}\cdot\hat{\Omega}/(v_{gw}/c))\right]\times\right.\\
&\exp\left[-i\frac{f}{2f^*}(3+\hat{n}\cdot\hat{\Omega}/(v_{gw}/c))\right] \\
& +\text{sinc}\left[\frac{f}{2f^*}(1+\hat{n}\cdot\hat{\Omega}/(v_{gw}/c))\right]\times\\
&\left.\exp\left[-i\frac{f}{2f^*}(1+\hat{n}\cdot\hat{\Omega}/(v_{gw}/c))\right]\right\},
\end{split}
\end{equation}
where $L$ is the arm length of the detector and $f^*=c/(2\pi L)$ is the transfer frequency. Given that the sources come from all directions, we take the average of all directions $(\theta,\phi)$ and the polarization angle $\psi$ to obtain the averaged response (transfer) function, as follows:
\begin{equation}
R^A=\frac{1}{8\pi^2}\int_0^\pi\sin\theta d\theta\int_0^{2\pi}d\phi\int_0^{2\pi}d\psi\left|F^A\right|^2.
\end{equation}
In this study, we set $G=c=h=1$.

\subsection{Tensor mode}
\begin{figure}[htp]
\centerline{\includegraphics[width=0.9\columnwidth]{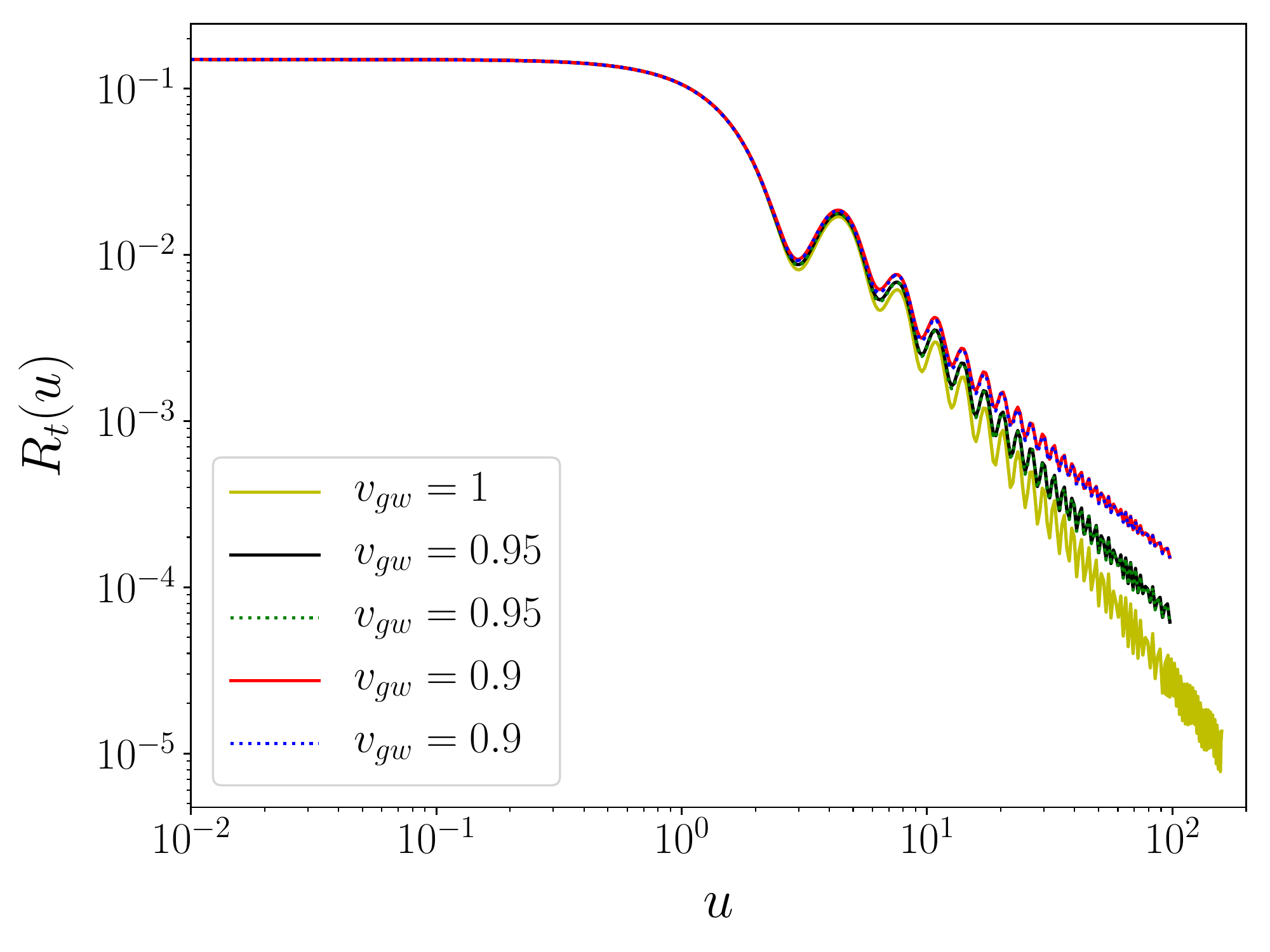}}
\caption{Averaged response functions of tensor modes obtained using the analytical approximations and semi-analytical formulas for interferometric gravitational wave (GW) detectors without optical cavities in the arms. We choose $\gamma=\pi/3$. The solid lines denote the semi-analytical formulas, and the dashed lines denote the analytical approximations.}
\label{fig1}
\end{figure}

The averaged angular response function for tensor modes is expressed as follows \cite{Larson:1999we,Liang:2019pry,Zhang:2019oet}:
\begin{equation}
\label{gqrt1}
\begin{split}
R^{+}(u)=& R^{\times}(u)= H(u)-\frac{1}{16\pi u^{2}}\int_{0}^{2\pi}d\epsilon\int_{0}^{\pi}d\theta_{1}\sin^3(\theta_1)\\
&\qquad\qquad \times \left[\sin^2(\theta_2)-2\sin^2(\gamma)
\sin^2(\epsilon)\right]\eta(u),
\end{split}
\end{equation}
where $H(u)$ is given in Eq. \eqref{hu} in the Appendix,
\begin{equation}
\label{gqeta1}
\begin{split}
\eta(u)=&\{[\cos(u)-\cos(u\mu_1)][\cos(u)-\cos(u\mu_2)]\mu_1\mu_2 \\
&+[\sin(u)-\mu_1\sin(u\mu_1)][\sin(u)-\mu_2\sin(u\mu_2)]\}/\\
&[(1-\mu_1^2)(1-\mu_2^2)],
\end{split}
\end{equation}
$u=2\pi f L/c$, $\mu_1=\cos\theta_1/v_{gw}$, $\mu_2=\cos\theta_{2}/v_{gw}$, $\cos\theta_2=\cos\gamma\cos\theta_1+\sin\gamma\sin\theta_1\cos\epsilon$, and $\gamma$ is the opening angle between two arms. An analytical expression for the integration in Eq. \eqref{gqrt1} is still unattainable, but we can calculate the integration numerically. We show the numerical result of $R^{+}(u)=R^{\times}(u)$ for different $v_{gw}$ in Fig. \ref{fig1}. For massless GWs, $v_{gw}=c$, the full analytical expression for $R^{+}(u)=R^{\times}(u)=R_{t}(u)/2$ was presented in Ref. \cite{Zhang:2020khm}, and the result is given in Eq. \eqref{anatensor} in the Appendix.
In the low-frequency limit, $u\rightarrow 0$, $\eta(u)$ becomes $\eta(u) = u^2$, and the integration in Eq. \eqref{gqrt1} is rewritten as follows:
\begin{equation}
\label{gqrt4}
\begin{split}
&\frac{1}{16\pi u^{2}}\int_{0}^{2\pi}d\epsilon\int_{0}^{\pi}d\theta_{1}\sin^3(\theta_1) \times \\
&\qquad \left[\sin^2(\theta_2)-2\sin^2(\gamma)\sin^2(\epsilon)\right]\eta(u)\\
&=[1+3\cos(2\gamma)]/30.
\end{split}
\end{equation}
This result is the same as that of the massless case with $v_{gw}=c$ in the low-frequency limit. Therefore, we propose to approximate the integration in Eq. \eqref{gqrt1} with the analytical result obtained in GR in which GWs propagate with the speed of light, i.e., the analytical approximation for the averaged response function for the tensor modes \eqref{gqrt1} is expressed as follows:
\begin{equation}
\label{gqrt2}
\begin{split}
R^{+}_{aa}(u)=R^{\times}_{aa}(u)&=H(u)+T(u),
\end{split}
\end{equation}
where the subscript $aa$ denotes analytical approximation and $T(u)$ is the integration in Eq. \eqref{gqrt1} with $v_{gw}$ being equal to the speed of light \cite{Zhang:2020khm}, and it is given in Eq. \eqref{mivt} in the Appendix. Using Eq. \eqref{gqrt2}, we plot the approximation of $R^{+}(u)=R^{\times}(u)$ for different $v_{gw}$ in Fig. \ref{fig1}. Fig. \ref{fig1} shows that the difference between the approximation and the numerical result is negligible even for $v_{gw} = 0.9$. Thus, the analytical expression in Eq. \eqref{gqrt2} approximates the averaged response function for the tensor modes very well.

\subsection{Vector mode}
\begin{figure}[htp]
\centerline{\includegraphics[width=0.9\columnwidth]{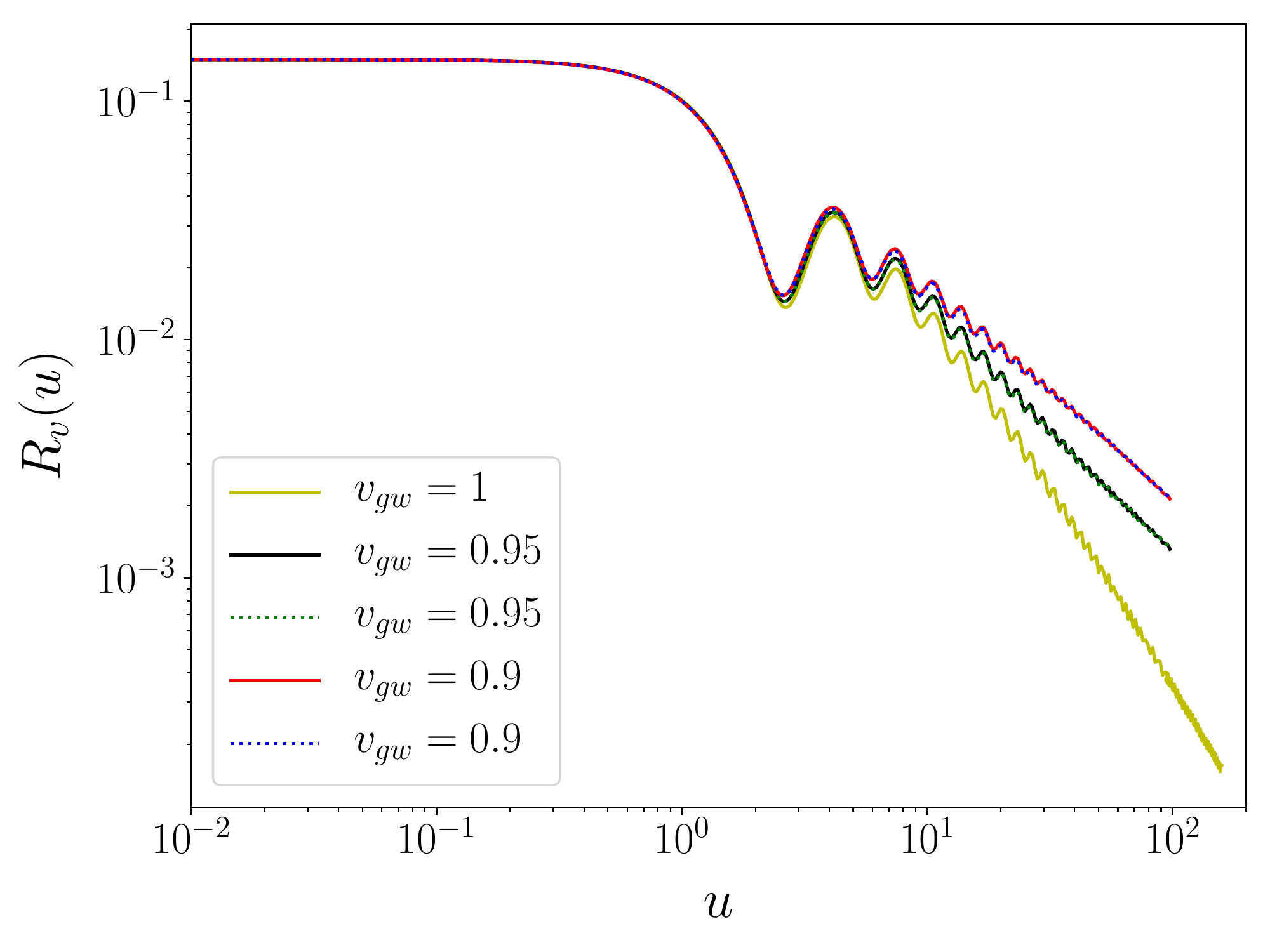}}
\caption{Averaged response functions of vector modes obtained using the analytical approximations and semi-analytical formulas for interferometric GW detectors without optical cavities in the arms. We choose $\gamma=\pi/3$. The solid lines denote the semi-analytical formulas, and the dashed lines denote the analytical approximations.}
\label{fig2}
\end{figure}

The semi-analytical formula of the averaged response function for the vector modes is expressed as follows \cite{Zhang:2019oet}:
\begin{equation}
\label{gqrv1}
\begin{split}
R^{x}(u)=&R^{y}(u)\\
=&A_v(u)-\frac{1}{8\pi u^{2}}\int_{0}^{2\pi}d\epsilon\int_{0}^{\pi}d\theta_1 \sin(\theta_{1})\sin(2\theta_{1})\cos(\theta_{2})\\
&\times \left[\cos(\gamma)\sin(\theta_{1})-\sin(\gamma)\cos(\theta_{1})\cos(\epsilon)\right]\eta(u),
\end{split}
\end{equation}
where $A_v(u)$ is given in Eq. \eqref{gqrv2} in the Appendix. We cannot integrate the integration in Eq. \eqref{gqrv1} to obtain an analytical expression. However, we can perform the integration numerically. We show the numerical result of $R^{x}(u)=R^{y}(u)$ for different $v_{gw}$ in Fig. \ref{fig2}. Based on the same argument, we approximate the averaged response function for the vector modes as follows:
\begin{equation}
\label{rx}
R^{x}_{aa}(u)=R^{y}_{aa}(u)
=A_v(u)+V(u),
\end{equation}
where $V(u)$ is given in Eq. \eqref{mivv} in the Appendix.

Using Eq. \eqref{rx}, we plot the approximation of $R^{x}(u)=R^{y}(u)$ for different $v_{gw}$ in Fig. \ref{fig2}. Fig. \ref{fig2} shows that the difference between the approximation and the numerical result is negligible even for $v_{gw} = 0.9$. Thus, the analytical expression in Eq. \eqref{gqrt2} approximates the averaged response function for the vector modes very well.

\subsection{Breathing mode}
\begin{figure}[htp]
\centerline{\includegraphics[width=0.9\columnwidth]{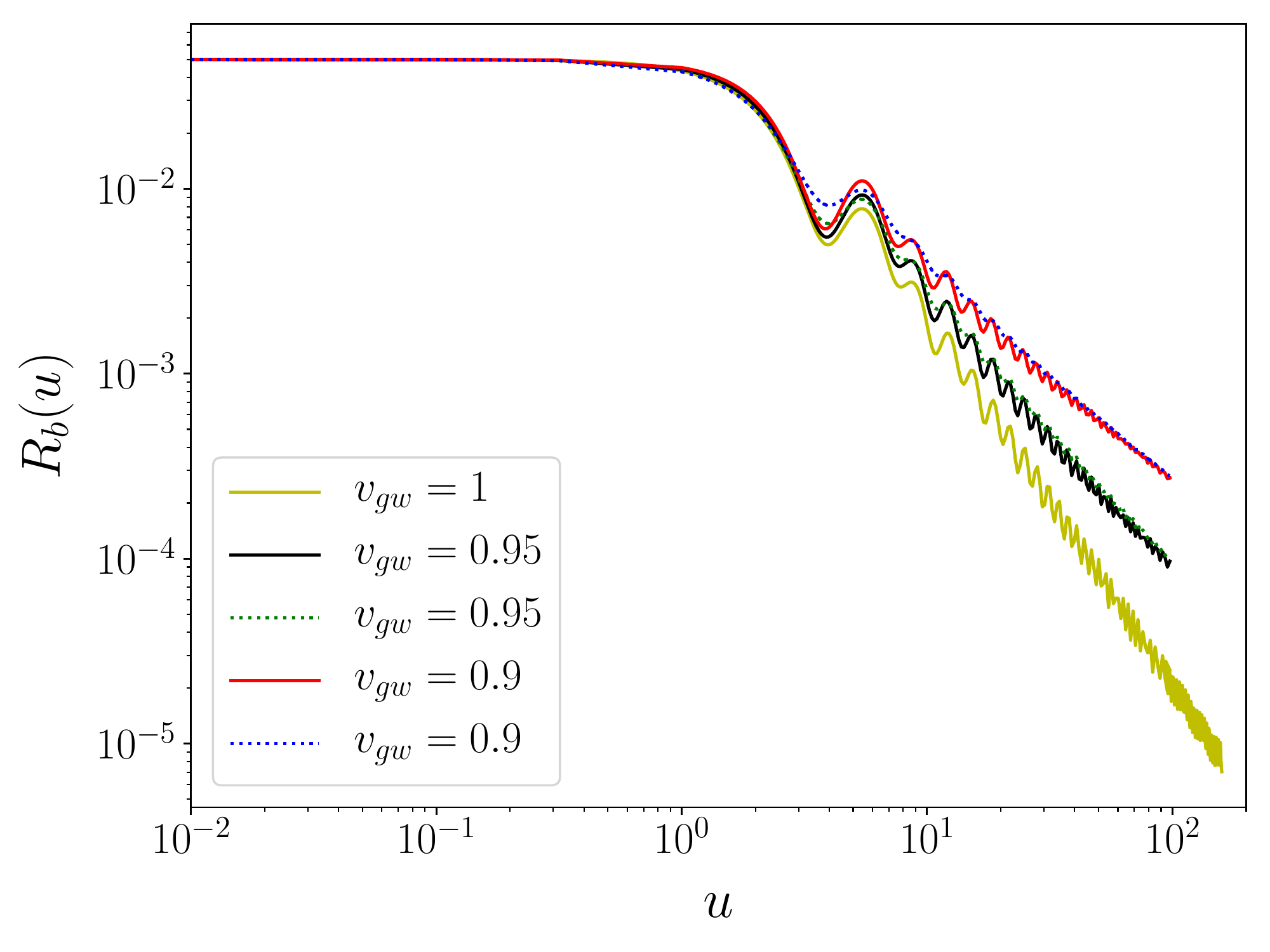}}
\caption{Averaged response functions of breathing modes obtained using the analytical approximations and semi-analytical formulas for interferometric GW detectors without optical cavities in the arms. We choose $\gamma=\pi/3$. The solid lines denote the semi-analytical formulas, and the dashed lines denote the analytical approximations.}
\label{fig3}
\end{figure}

The semi-analytical formula of the averaged response function for the breathing modes is expressed as follows \cite{Zhang:2019oet}:
\begin{equation}
\label{rb}
R^b(u)=2H(u)-\frac{1}{8\pi u^{2}}\int_{0}^{2\pi}d\epsilon\int_{0}^{\pi}d\theta_{1}\sin^3(\theta_{1})\sin^2(\theta_{2})\eta(u).
\end{equation}
We cannot integrate the integration in Eq. \eqref{rb} to obtain an analytical expression. However, we can perform the integration numerically. We show the numerical result of $R^{b}(u)$ for different $v_{gw}$ in Fig. \ref{fig3}. We approximate the averaged response function for the breathing modes as follows:
\begin{equation}
\label{gqrb}
R^b_{aa}(u)=2H(u)+B(u),
\end{equation}
where $B(u)$ is given in Eq. \eqref{mib} in the Appendix. Using Eq. \eqref{gqrb}, we plot the approximation of $R^{b}(u)$ for different $v_{gw}$ in Fig. \ref{fig3}. Fig. \ref{fig3} shows that the difference between the approximation and the numerical result is negligible even for $v_{gw} = 0.9$. Thus, the analytical expression \eqref{gqrb} approximates the averaged response function for the breathing modes very well.

\subsection{Longitudinal mode}
\begin{figure}[htp]
\centerline{\includegraphics[width=0.9\columnwidth]{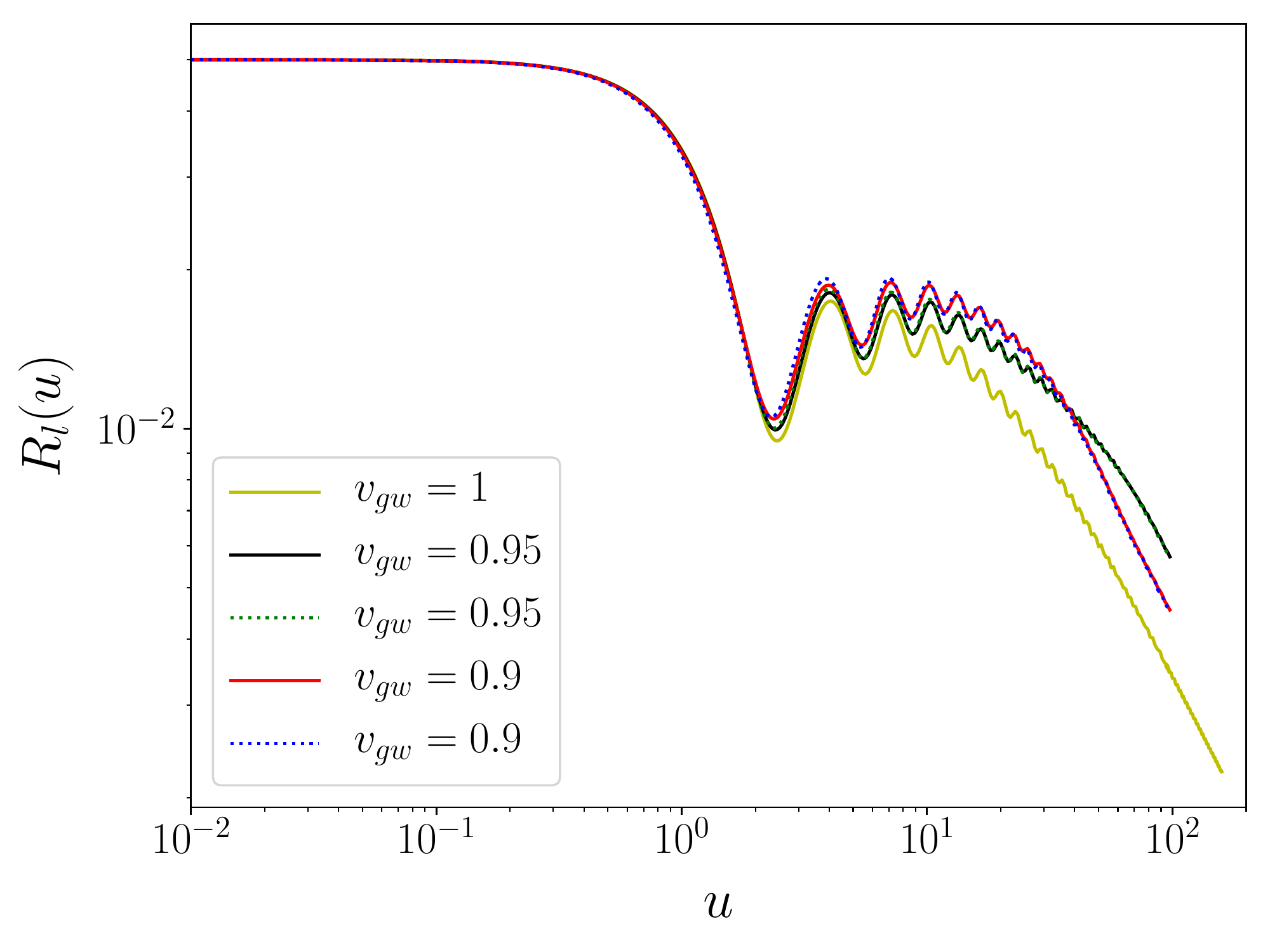}}
\caption{Averaged response functions of longitudinal modes obtained using the analytical approximations and semi-analytical formulas for interferometric GW detectors without optical cavities in the arms. We choose $\gamma=\pi/3$. The solid lines denote the semi-analytical formulas, and the dashed lines denote the analytical approximations.}
\label{fig4}
\end{figure}

The semi-analytical formula of the averaged response function for the longitudinal modes is expressed as follows \cite{Zhang:2019oet}:
\begin{equation}
\label{rl}
\begin{split}
R^l(u)=&A_l(u)-\frac{1}{8\pi u^{2}}\int_{0}^{2\pi}d\epsilon\int_{0}^{\pi}
d\theta_{1}\sin(\theta_{1})\times \\
&\qquad\qquad \cos^2(\theta_{1})\cos^2(\theta_{2})\eta(u),
\end{split}
\end{equation}
where $A_l(u)$ is given in Eq. \eqref{rl1} in the Appendix. We cannot integrate the integration in Eq. \eqref{rl} to obtain an analytical expression. However, we can perform the integration numerically. We show the numerical result of $R^{l}(u)$ for different $v_{gw}$ in Fig. \ref{fig4}.
The averaged response function for the longitudinal modes is approximated as follows:
\begin{equation}
\label{gqrl}
R^l_{aa}(u)=A_l(u)+L(u),
\end{equation}
where $L(u)$ is given in Eq. \eqref{mil} in the Appendix. Using Eq. \eqref{gqrl}, we plot the approximation of $R^{l}(u)$ for different $v_{gw}$ in Fig. \ref{fig4}. Fig. \ref{fig4} shows that the difference between the approximation and the numerical result is negligible even for $v_{gw}=0.9$. Thus, the analytical expression \eqref{gqrb} approximates the averaged response function for the longitudinal modes very well.

\section{Constraint on the mass of graviton}
\label{sec3}
Now, we consider the parameter estimation and the constraint on the mass of graviton. The noise spectral density $S_n(f)$ for space-based GW detectors is expressed as follows:
\begin{equation}
S_n(f)=\frac{1}{L^2}\left\{S_x+\left[1+\left(\frac{0.4\text{mHz}}{f}\right)^2\right]\frac{4S_a}{(2\pi f)^4}\right\}+S_c(f),
\end{equation}
where the confusion noise is
\begin{equation}
S_c(f)=A f^{-7/3}\exp[-f^\alpha+\beta f \sin(\kappa f)][1+\tanh (\gamma(f_\kappa-f))]\text{Hz}^{-1}
\end{equation}
and the arm length $L$, position noise $\sqrt{S_x}$, and acceleration noise $\sqrt{S_a}$ are the parameters of the detector. $L = 2.5\times 10^9$ m, $\sqrt{S_x} = 15\ \text{pm/Hz}^{1/2}$, and $\sqrt{S_a} = 3\times10^{-15}\ \text{m s}^{-2}/\text{Hz}^{1/2}$ for LISA \cite{Danzmann:1997hm,LISA:2017pwj}; $L=\sqrt{3}\times 10^8$ m, $\sqrt{S_x} = 1\ \text{pm/Hz}^{1/2}$, and $\sqrt{S_a} = 10^{-15}\ \text{m s}^{-2}/\text{Hz}^{1/2}$ for TianQin \cite{TianQin:2015yph}; and $L = 3\times 10^9$ m, $\sqrt{S_x} = 8\ \text{pm/Hz}^{1/2}$, and $\sqrt{S_a} = 3\times10^{-15}\ \text{m s}^{-2}/\text{Hz}^{1/2}$ for Taiji \cite{Hu:2017mde,Ruan:2018tsw}. With a given noise spectral density $S_n(f)$ for the detector, the inner product between two signals $h_1(t)$ and $h_2(t)$ is defined as follows:
\begin{equation}
\label{gqprod1}
(h_1|h_2)\equiv2\int_0^\infty\frac{\tilde{h_1^*}\tilde{h_2}+\tilde{h_2^*}\tilde{h_1}}{S_n(f)}df,
\end{equation}
where $\tilde{h_1}(f)$ and $\tilde{h_2}(f)$ are the Fourier transforms of the respective gravitational waveforms $h(t)$ registered in the detector. The signal-to-noise ratio (SNR) for a given $h$ is derived as follows:
\begin{equation}
\rho[h]\equiv(h|h)^{1/2}.
\end{equation}
The waveform is characterized by a set of parameters $\theta^a$, and the parameter error with a large SNR limit is estimated as follows:
\begin{equation}
\Delta\theta^a=\sqrt{\langle (\theta^a)^2-\langle \theta^a \rangle^2\rangle}\approx \sqrt{\Sigma_{aa}},
\end{equation}
where $\Sigma_{aa}$ is the diagonal element of the inverse of the FIM $\Sigma_{ab}=(\Gamma^{-1})_{ab}$. The FIM $\Gamma$ is defined as follows:
\begin{equation}
\Gamma_{ab}\equiv\left(\frac{\partial h}{\partial\theta^a}\left|\frac{\partial h}{\partial\theta^b}\right.\right).
\end{equation}
The covariant matrix between two parameters $\theta^a$ and $\theta^b$ is expressed as follows:
\begin{equation}
C_{ab}=\Sigma_{ab}/\sqrt{\Sigma_{aa}\,\Sigma_{bb}}.
\end{equation}
To obtain the parameter estimation, we use the frequency domain waveform, as follows \cite{Huerta:2014eca,Yagi:2009zm,Pratten:2020fqn}:
\begin{equation}
\tilde{h}(f)=\frac{\sqrt{3}}{2}\mathcal{A}f^{-7/6}e^{i\Psi(f)},
\end{equation}
where the amplitude with 3 PN order is derived as follows:
\begin{equation}
\begin{split}
\mathcal{A}=&\frac{1}{\sqrt{30}\pi^{2/3}}\frac{\mathcal{M}^{5/6}}{D_L}\left\{1+\left(-\frac{323}{224}+\frac{451\eta}{168}\right)(\pi M f)^{2/3}\right.\\
&+\left(\frac{105271}{24192}\eta^2-\frac{1975055}{338688}\eta
-\frac{27312085}{8128512}\right)(\pi M f)^{4/3}\\
&+\left(-\frac{85\pi}{64}+\frac{85\pi\eta}{16}\right)(\pi M f)^{5/3}+\left[-\frac{177520268561}{8583708672}\right.\\
&+\left(\frac{545384828789}{5007163392}-\frac{205\pi^2}{48}\right)\eta-\frac{3248849057\eta^2}
{178827264}\\
&\left.\left.+\frac{34473079\eta^3}{6386688}\right](\pi M f)^{2}\right\}
\end{split}
\end{equation}
and the phase with 3.5 PN order is expressed as follows:
\begin{equation}
\begin{split}
\Psi(f)=&2\pi f t_c-\phi_c-\frac{\pi}{4}+\frac{3}{128\eta}(\pi M f)^{-5/3}\left\{1\right.\\
&+\left(\frac{3715}{756}+\frac{55\eta}{9}\right)(\pi M f)^{2/3}
-\frac{128}{3}\frac{\pi^2D\eta M}{\lambda_g^2(1+z)}(\pi M f)^{2/3}\\
&-16\pi(\pi M f)+\left(\frac{15293365}{508032}+\frac{27145\eta}{504}\right.\\
&\left.+\frac{3085\eta^2}{72}\right)(\pi M f)^{4/3}+\left(\frac{-65\pi\eta}{9}\right.\\
&-\frac{65}{3}\pi\eta \log[\sqrt{6}(\pi M f)^{1/3}]+\frac{38645}{252}\pi \log[\sqrt{6}(\pi M f)^{1/3}]\\
&\left.+\frac{38645\pi}{756}\right)(\pi M f)^{5/3}
+\left(\frac{-127825\eta^3}{1296}\right.\\
&+\frac{76055\eta^2}{1728}-\frac{6848\gamma_E}{21}+\frac{2255\pi^2\eta}{12}
-\frac{15737765635\eta}{3048192}\\
&-\frac{6848}{21}\log[4(\pi M f)^{1/3}]
-\frac{640\pi^2}{3}\\
&\left.+\frac{11583231236531}{4694215680}\right)(\pi M f)^{2}+\left(-\frac{74045\pi\eta^2}{756}\right.\\
&\left.\left.+\frac{378515\pi\eta}{1512}+\frac{77096675\pi}{254016}\right)(\pi M f)^{7/3}\right\},
\end{split}
\end{equation}
where $f$ is the frequency of the GW, $\mathcal{M}=\eta^{3/5}M$ is the chirp mass (where $M=m_1+m_2$, $\eta=m_1m_2/M^2$), and $D_L$ is the luminosity distance to the source derived as follows:
\begin{equation}
D_L=\frac{1+z}{H_0}\int_{0}^{z}\frac{dz'}{[\Omega_M(1+z')^3+\Omega_\Lambda]^{1/2}}.
\end{equation}
The quantity $D$ is expressed as follows:
\begin{equation}
D=\frac{1+z}{H_0}\int_{0}^{z}\frac{dz'}{(1+z')^2[\Omega_M(1+z')^3+\Omega_\Lambda]^{1/2}},
\end{equation}
where $H_0$ is the Hubble constant, $\Omega_M$ is the current matter energy density, and $\Omega_\Lambda$ is the current energy density of dark energy, which is taken as the cosmological constant for simplicity. We use the Planck 2018 constraints $H_0 = 67.4\ \mathrm{kms}^{-1}\mathrm{Mpc}^{-1}$, $\Omega_M = 0.315$, and $\Omega_\Lambda = 0.685$ \cite{Planck:2018jri}.
To identify the waveforms in the detector, we do not consider the effects of the directions of the sources, i.e., we consider the averaged response function for the source. Thus, for the inner product in Eq. \eqref{gqprod1}, we need to replace $S_n(f)$ with $S_n(f)/R^A(f)$. In this study, we consider the tensor modes only. To easily calculate the FIM, we should use an analytical expression for the averaged response function. As discussed in the previous section, there is no analytical expression for the averaged response function in the massive case. However, we can use the approximate analytical results instead. To determine the effect of the response function for massive GWs, we compare the results of the errors of the parameters using both Eqs. \eqref{anatensor} and \eqref{gqrt2} for LISA, and the results are shown in Table \ref{table1}. We set $D_L = 3\ $Gpc and consider the 1-year observation of equal-mass massive BBHs with different masses obtained using LISA in Table \ref{table1}. Table \ref{table1} shows that the results are nearly the same for the response function with and without the component mass. Therefore, the effect of the mass of graviton on the response function can be neglected, and we can use the full analytical expression in Eq. \eqref{anatensor} to conduct parameter estimation. In the subsequent discussion, we use the full analytical expression in Eq. \eqref{anatensor} to estimate the mass of graviton.

\begin{table*}[htbp]
	\renewcommand\tabcolsep{4.0pt}
	\begin{tabular}{llllllllll}
		\hline
		\hline
		Binaries  & $R_{aa}$ & SNR  &$\Delta \mathrm{ln}\mathcal{M}$& $\Delta \mathrm{ln}\eta$ &$\Delta \mathrm{ln}D_L$ &$\Delta t_c\ (\text{s})$ &$\Delta\phi_c$&$ m_g$&$\lambda $\\
		($M_{\odot}$) &  &  &($10^{-5}$)&  & & &&($10^{-61}\text{ kg}$)&($10^{17}\text{ m}$)\\
		\hline
 		$10^3/10^3$ & $\times$  &$15.17$~& $ 1.747$ ~& $0.0249$~&$0.0659$~&$5.052$~&$3.827$~&$25.6$~&$8.64$\\
        ~ & $\surd$  &$15.17$~& $ 1.764$ ~& $0.0252$~&$0.0659$~&$5.158$~&$3.887$~&$25.7$~&$8.60$\\
        $10^4/10^4$ & $\times$  &$105.56$~& $ 1.722$ ~& $0.0052$~&$0.0095$~&$0.750$~&$0.206$~&$6.19$~&$35.7$\\
        ~ & $\surd$  &$105.56$~& $ 1.722$ ~& $0.0052$~&$0.0095$~&$0.750$~&$0.206$~&$6.19$~&$35.7$\\
        $10^5/10^5$ & $\times$  &$656.63$~& $ 2.350$ ~& $0.0018$~&$0.0015$~&$0.349$~&$0.012$~&$1.63$~&$136$\\
        ~ & $\surd$  &$656.63$~& $ 2.350$ ~& $0.0018$~&$0.0015$~&$0.348$~&$0.012$~&$1.63$~&$136$\\
		\hline
		\hline
	\end{tabular}
	\caption{Results of parameter errors for different equal-mass BBHs in massive gravity theory with 1-year observation before the ISCO obtained using LISA. “$\surd$” means that we use the approximate analytical expression in Eq. \eqref{gqrt2} of the averaged response function for the massive case, while “$\times$” means that we use the full analytical expression in Eq. \eqref{anatensor} of the averaged response function for the massless case with $m_g = 0$.}
\label{table1}
\end{table*}

Now, we compare the parameter estimations for equal-mass BBHs with LISA, Taiji, and TianQin \cite{Gong:2021gvw,Zhang:2021wwd}. We set the luminosity distance of the binaries at $D_L = 3\ $Gpc and start the observation one year before the coalescence. Thus, the upper limit of the integral in Eq. \eqref{gqprod1} is the frequency at the innermost stable circular orbit (ISCO), $f_\text{ISCO} = (6\pi\sqrt{6} GM)^{-1}$. The mass of massive and supermassive BBHs is in the range $M = 10^3-10^6M_{\odot}$, and the results are shown in Table \ref{table2}. The results showed that Taiji has the largest SNR and provides the strongest constraint on the mass of graviton for all of the binaries considered. If the mass of BBHs increases, then the frequency of GWs decreases. When the mass reaches $10^6M_\odot$, some GW signals are out of the sensitive band of space-based GW detectors, and as the SNR in the detector decreases, the errors of the parameters, except for the mass of graviton, increase. When the mass of BBHs is smaller than $10^4\ M_{\odot}$, TianQin provides the smallest $\Delta t_c$ and $\Delta\phi_c$. By contrast, when the mass of BBHs is larger than $10^4\ M_{\odot}$, Taiji provides better constraints on the parameters. The wavelength of massive graviton can be constrained to be $\lambda_g > 1.91\times 10^{19}$ m and the mass can be constrained to be $m_g < 1.16\times 10^{-61}$ kg by Taiji.

\begin{table*}[htbp]
	\renewcommand\tabcolsep{4.0pt}
	\begin{tabular}{llllllllll}
		\hline
		\hline
			Binaries & Detectors & SNR  &$\Delta\ln\mathcal{M}$ & $\Delta \ln\eta$ & $\Delta \ln D_L$ & $\Delta t_c(\text{s})$ & $\Delta\phi_c$ & $m_g$ & $\lambda$\\
		($M_{\odot}$) &  &  &($10^{-6}$)&  & & &&($10^{-61}\text{ kg}$)&($10^{17}\text{ m}$)\\
		\hline
 		$10^3/10^3$ & LISA  &$15.17$~& $ 17.47$ ~& $0.0249$~&$0.0659$~&$5.052$~&$3.827$~&$25.6$~&$8.64$\\
        ~ & Taiji  &$29.64$~& $ 9.724$ ~& $0.0136$~&$0.0337$~&$2.683$~&$2.062$~&$19.0$~&$1.16$\\
        ~ & TianQin  &$11.94$~& $ 13.48$ ~& $0.0135$~&$0.0084$~&$0.908$~&$1.442$~&$22.0$~&$10.1$\\
        $10^4/10^4$ & LISA  &$105.56$~& $ 17.22$ ~& $0.0052$~&$0.0095$~&$0.750$~&$0.206$~&$6.19$~&$35.7$\\
        ~ & Taiji  &$201.53$~& $ 12.05$ ~& $0.0032$~&$0.0050$~&$0.408$~&$0.115$~&$5.00$~&$44.2$\\
        ~ & TianQin  &$80.72$~& $ 17.34$ ~& $0.0040$~&$0.0124$~&$0.262$~&$0.105$~&$5.88$~&$37.6$\\
        $10^5/10^5$ & LISA  &$656.63$~& $ 23.50$ ~& $0.0018$~&$0.0015$~&$0.349$~&$0.012$~&$1.63$~&$136$\\
        ~ & Taiji  &$1242.49$~& $ 19.45$ ~& $0.0013$~&$0.0008$~&$0.212$~&$0.010$~&$1.45$~&$153$\\
        ~ & TianQin  &$450.25$~& $ 53.30$ ~& $0.0031$~&$0.0022$~&$0.479$~&$0.024$~&$2.27$~&$97.6$\\
        $10^6/10^6$ & LISA  &$337.90$~& $ 219.3$ ~& $0.0074$~&$0.0031$~&$8.261$~&$0.085$~&$1.22$~&$182$\\
        ~ & Taiji  &$343.59$~& $ 195.0$ ~& $0.0068$~&$0.0030$~&$7.785$~&$0.076$~&$1.16$~&$191$\\
        ~ & TianQin  &$239.97$~& $ 699.3$ ~& $0.0181$~&$0.0048$~&$16.90$~&$0.259$~&$2.02$~&$110$\\
		\hline
		\hline
	\end{tabular}
	\caption{Results of parameter errors for different equal-mass BBHs in massive gravity theory with 1-year observation before the ISCO obtained using LISA, Taiji, and TianQin. We use the full analytical expression in Eq. \eqref{anatensor} for the averaged response function.}
\label{table2}
\end{table*}

Intermediate/extreme mass ratio inspirals (IMRIs/EMRIs) are one of the most promising GW sources for space-based GW detectors. The mass ratio is approximately $10^2$-$10^4:1$ for IMRIs and $\gtrsim 10^4:1$ for EMRIs.
We also consider IMRIs/EMRIs consisting of a small compact object (CO) with the mass of $10M_{\odot}$ and a BH with the mass $M_{BH} = 400M_{\odot}$, $10^3M_{\odot}$, $10^4M_{\odot}$, $10^5M_{\odot}$
in massive gravity theory.
We fix SNR = 10 for all IMRI/EMRIs with different detectors. The observation time is one year before the ISCO, and the results of parameter errors and the constraints on the mass of graviton obtained using LISA, Taiji, and TianQin are summarized in Table \ref{table3}.
For the IMRIs/EMRIs that we considered, Taiji provides a better constraint on the mass of graviton than LISA.
When the mass of the central BH in the IMRIs is equal to or less than $10^4M_\odot$, TianQin provides the best constraint on the mass of graviton because TianQin is more sensitive in the relatively high-frequency bands.
When the mass of the central BH in the IMRIs reaches $10^5M_\odot$, Taiji provides the best constraint on the mass of graviton.
The mass of graviton can be constrained to be $m_g<1.04\times 10^{-59}$ kg, and the wavelength of massive graviton can be constrained to be $\lambda_g > 2.13\times 10^{17}$ m for the $10M_\odot$/$10^4M_\odot$ IMRIs by TianQin.

\begin{table*}[htbp]
\renewcommand\tabcolsep{4.0pt}
	\begin{tabular}{llllllllll}
		\hline
		\hline
			Binaries & Detectors & $z$  &$\Delta \ln\mathcal{M}$& $\Delta \ln\eta$ &$\Delta \ln D_L$ &$\Delta t_c(\text{s})$ &$\Delta\phi_c$&$m_g$&$\lambda$\\
			($M_{\odot}$) &  &  &($10^{-6}$)&  & & &&($10^{-59}\text{ kg}$)&($10^{17}\text{ m}$)\\
		\hline
 		$10/400$ & LISA  &$0.0146$~& $ 6.995$ ~& $0.0009$~&$0.0994$~&$3.789$~&$14.45$~&$2.54$~&$0.87$\\
        ~ & Taiji  &$0.0275$~& $ 6.864$ ~& $0.0093$~&$0.0998$~&$3.855$~&$14.58$~&$1.89$~&$1.17$\\
        ~ & TianQin  &$0.028$~& $ 3.331$ ~& $0.0035$~&$0.0999$~&$0.770$~&$4.391$~&$1.07$~&$2.06$\\
        $10/10^3$ & LISA  &$0.027$~& $ 8.566$ ~& $0.0059$~&$0.0991$~&$4.699$~&$13.73$~&$1.91$~&$1.16$\\
        ~ & Taiji  &$0.0528$~& $ 9.064$ ~& $0.0063$~&$0.0998$~&$4.979$~&$14.63$~&$1.41$~&$1.56$\\
        ~ & TianQin  &$0.0438$~& $ 4.165$ ~& $0.0023$~&$0.0997$~&$0.921$~&$4.203$~&$1.22$~&$1.81$\\
        $10/5000$ & LISA  &$0.0705$~& $ 13.11$ ~& $0.0029$~&$0.0994$~&$7.441$~&$13.85$~&$1.75$~&$1.26$\\
        ~ & Taiji  &$0.141$~& $ 13.55$ ~& $0.0030$~&$0.0999$~&$8.003$~&$14.47$~&$1.31$~&$1.69$\\
        ~ & TianQin  &$0.093$~& $ 7.098$ ~& $0.0013$~&$0.0999$~&$1.659$~&$4.731$~&$1.13$~&$1.96$\\
        $10/10^4$ & LISA  &$0.102$~& $ 16.85$ ~& $0.0023$~&$0.0992$~&$9.638$~&$14.70$~&$1.58$~&$1.40$\\
        ~ & Taiji  &$0.203$~& $ 17.25$ ~& $0.0024$~&$0.0994$~&$10.50$~&$15.47$~&$1.20$~&$1.84$\\
        ~ & TianQin  &$0.125$~& $ 9.129$ ~& $0.0010$~&$0.0996$~&$2.282$~&$5.133$~&$1.04$~&$2.13$\\
        $10/10^5$ & LISA  &$0.293$~& $ 66.62$ ~& $0.0022$~&$0.0998$~&$54.11$~&$42.54$~&$1.49$~&$1.49$\\
        ~ & Taiji &$0.552$~& $ 71.42$ ~& $0.0024$~&$0.0995$~&$62.00$~&$46.75$~&$1.27$~&$1.75$\\
        ~ & TianQin  &$0.269$~& $ 56.24$ ~& $0.0018$~&$0.0996$~&$33.56$~&$31.71$~&$1.38$~&$1.60$\\
		\hline
		\hline
	\end{tabular}
	\caption{Results of parameter errors for different IMRIs/EMRIs in massive gravity theory with 1-year observation before the ISCO obtained using LISA, Taiji, and TianQin. We set SNR $\rho = 10$ and use the full analytical expression in Eq. \eqref{anatensor} for the averaged response function. The mass of the small CO is set at $10M_\odot$.}
\label{table3}
\end{table*}

\section{Discussion and Conclusions}
\label{sec4}
To calculate the parameter errors using the FIM method, we need to utilize the averaged response function. However, the propagation speed $v_{gw}$ of GWs is different from the speed of light $c$ in massive gravity theory, and analytical expressions of the averaged response functions for GWs propagating with speed $v_{gw}\neq c$ are unavailable. We note that the remaining integral in the averaged response function for GWs propagating with speed $v_{gw}\neq c$ is the same as that for GWs propagating with $c$ in the low-frequency limit. Thus, we propose approximating the averaged response functions for GWs propagating with speed $v_{gw}\neq c$ by replacing the remaining integral with those for GWs propagating with $c$. We compare the analytical approximations with the numerical results of the averaged response functions for GWs propagating with speed $v_{gw}\neq c$ and determine that the analytical expressions approximate the numerical results well. Then, we consider the effect of the averaged response function for tensor modes on parameter estimation using LISA.
We compare the parameter errors using the analytical approximation of the averaged response function for GWs propagating with speed $v_{gw}$ and the analytical result of the averaged response function for GWs propagating with speed $c$ and determine that the effect of the response function on the parameter estimation is negligible for small mass $m_g$.
Therefore, during parameter estimation, we do not need to consider the mass of graviton in the averaged response function, and we can use the averaged response function for GWs propagating with speed $c$.
Using the FIM method and the 3.5 PN waveform template, we perform parameter estimations using LISA, Taiji, and TianQin for both equal-mass BBHs and IMRIs/EMRIs in massive gravity theory.
The observation period is one year before the ISCO. The results showed that Taiji has the largest SNR and provides the strongest constraint on the mass of graviton for equal-mass BBHs. When the mass of BBHs is smaller than $10^4\ M_{\odot}$, TianQin has the smallest $\Delta t_c$ and $\Delta\phi_c$.
By contrast, when the mass of BBHs is larger than $10^4\ M_{\odot}$, Taiji provides better constraints on the parameters.
The wavelength of massive graviton can be constrained to be $\lambda_g > 1.91\times 10^{19}$ m, and the mass can be constrained to be $m_g < 1.16\times 10^{-61}$ kg by Taiji.
For the IMRIs/EMRIs, Taiji provides a better constraint on the mass of graviton than LISA.
When the mass of the central BH in the IMRIs/EMRIs is equal to or less than $10^4M_\odot$, TianQin provides the best constraint on the mass of graviton.
When the mass of the central BH in the IMRIs/EMRIs reaches $10^5M_\odot$, Taiji provides the best constraint on the mass of graviton.
In conclusion, the wavelength of massive graviton can be constrained to be $\lambda_g > 1.91\times 10^{19}$ m, and the mass can be constrained to be $m_g < 1.16\times 10^{-61}$ kg by 1-year observation of equal-mass BBHs.

\begin{acknowledgments}
This work is supported by the National Key Research \& Development Program of China (Grant No. 2020YFC2201504)
and the National Natural Science Foundation of China under Grant Nos. 12175184.
The author thanks Dr. Chao Zhang from Huazhong University of Science and Technology for helpful discussions.
\end{acknowledgments}



\appendix
\section{The analytical formulae}

\begin{equation}
\label{hu}
\begin{split}
H(u)=&\frac{-v_{gw}^2}{24 u^5}\left\{3 u^3 \left(7-9 v_{gw}^2\right)
+u^3 \left(5-3 v_{gw}^2\right) \cos(2u)
-12 u^2 v_{gw}^2 \sin (u) \cos \left(\frac{u}{v_{gw}}\right)
\right.\\
&+6u v_{gw} \sin (u) \left[\left(u^2+2\right) v_{gw}^2-u^2\right] \sin \left(\frac{u}{v_{gw}}\right)\\
&+6 \cos (u)\left(4v_{gw}^3 u^2-4v_{gw}^3 -2 v_{gw} u^2\right) \sin \left(\frac{u}{v_{gw}}\right)\\
&\left.+6 \cos (u)(u^3v_{gw}^2+4u v_{gw}^2-u^3) \cos \left(\frac{u}{v_{gw}}\right)\right\} \\
&+\frac{v_{gw}}{16 u^2} \left(v_{gw}^2-1\right) \left[ \left(v_{gw}^2-1\right) \cos (2 u)+9 v_{gw}^2-1\right]\\
&\qquad \qquad \times
\left[\text{Ci} \left(u+\frac{u}{v_{gw}}\right)-\text{Ci} \left(\frac{u}{v_{gw}}-u\right)+\ln \left(\frac{1-v_{gw}}{1+v_{gw}}\right)\right]\\
&+\frac{2 u+\sin (2 u)}{16 u^2}v_{gw} \left(v_{gw}^2-1\right)^2\left[\text{Si}\left(\frac{u}{v_{gw}}+ u\right)-\text{Si}\left(u-\frac{u}{v_{gw}}\right)\right],
\end{split}
\end{equation}

\begin{equation}
\label{anatensor}
\begin{split}
R_{t}(u)=&\frac{1}{u^2}\left\{\frac{3-\cos\gamma}{12}+\frac{-1+\cos\gamma }{u^2}
+2\sin^2\left(\frac{\gamma}{2}\right)
\left[\text{Ci}\left[2u\sin\left(\frac{\gamma}{2}\right)\right]
-\ln\left[\sin\left(\frac{\gamma}{2}\right)\right]
\right.\right.\\
&\left.-\text{Ci}(2u)\right]
+\frac{1+\csc^2\left(\frac{\gamma}{2}\right)}{8u^2}
\cos\left[2u\sin\left(\frac{\gamma}{2}\right)\right]
+\sin\left[2u\sin\left(\frac{\gamma}{2}\right)\right]
\left[\frac{-3+\cos\gamma}{32u^3}\right.\\
&\left.+\frac{-21+28\cos\gamma-7\cos(2\gamma)}{32u}\right]\csc^3\left(\frac{\gamma}{2}\right)
+\sin(2u)\left[\left(\frac{1}{u}+\frac{2}{u^3}\right)
\sin^2\left(\frac{\gamma}{2}\right)\right.\\
&\left.+\cos^2\left(\frac{\gamma}{2}\right)\Big(2\text{Si}(2u)-\text{Si}\left[2u+2u\sin\left(\frac{\gamma}{2}\right)\right]-\text{Si}\left[2u-2u\sin\left(\frac{\gamma}{2}\right)\right]\Big)\right]
\\
&+\left[\cos^2\left(\frac{\gamma}{2}\right)\left(2\text{Ci}(2u)+\ln\left[\cos^2\left(\frac{\gamma}{2}\right)\right]-\text{Ci}\left[2u+2u\sin\left(\frac{\gamma}{2}\right)\right]\right.\right.\\
&\left.\left.\left.-\text{Ci}\left[2u-2u\sin\left(\frac{\gamma}{2}\right)\right]\right)+\left(\frac{1}{6}-\frac{2}{u^2}\right)
\sin^2\left(\frac{\gamma}{2}\right)\right]\cos(2u)\right\}.
\end{split}
\end{equation}

\begin{equation}
\label{mivt}
T(u)=\frac{1}{2}R_t(u)-\frac{1}{4u^2}\left[[1+\cos^2(u)](\frac{1}{3}-\frac{2}{u^2}
)+\frac{2}{u^3}\sin(2u)+\sin^2(u)\right].
\end{equation}

\begin{equation}
\label{gqrv2}
\begin{split}
A_v(u)
=&\frac{v_{gw}^2}{12 u^5}\left\{
12 v_{gw}^2 \cos (u) \left[4 v_{gw} \left(u^2-1\right) \sin \left(\frac{u}{v_{gw}}\right)+u \left(u^2+4\right) \cos \left(\frac{u}{v_{gw}}\right)\right]\right.\\
&+2\left(2-3 v_{gw}^2\right) u^3 \cos(2u)+6 \left(2-9 v_{gw}^2\right) u^3\\
&\left.+2u\left[6 v_{gw}^3 \left(u^2+2\right) \sin (u) \sin \left(\frac{u}{v_{gw}}\right)-12 v_{gw}^2 u \sin (u) \cos \left(\frac{u}{v_{gw}}\right)\right]\right\}\\
&+\frac{v_{gw}^3}{2 u^2}\left(1-v_{gw}^2\right) \left[u+\sin (u) \cos (u)\right]\left[\text{Si}\left(u+\frac{u}{v_{gw}}\right)-\text{Si}\left(u-\frac{u}{v_{gw}}\right)\right]\\
&+\frac{v_{gw}^3}{4 u^2}\left[5-9v_{gw}^2+\left(1-v_{gw}^2\right)
\cos (2 u)\right]\left[\ln \left(\frac{1-v_{gw}}{1+v_{gw}}\right)+\text{Ci}\left(u+\frac{u}{v_{gw}}\right)-
\text{Ci}\left(\frac{u}{v_{gw}}-u\right)\right].
\end{split}
\end{equation}

\begin{equation}
\label{mivv}
\begin{split}
V(u)=&\frac{1}{2u^2}\left\{-4+\frac{4\cos\gamma}{3}+\frac{4-4\cos\gamma }{u^2}
+2\left[\gamma_E-\text{Ci}\left[2u\sin\left(\frac{\gamma}{2}\right)\right]
+\ln\left[2u\sin\left(\frac{\gamma}{2}\right)\right]\right]\right.\\
&-\frac{1+\csc^2\left(\frac{\gamma}{2}\right)}{2u^2}
\cos\left[2u\sin\left(\frac{\gamma}{2}\right)\right]
+\sin\left[2u\sin\left(\frac{\gamma}{2}\right)\right]
\left[\frac{7-8\cos\gamma+\cos(2\gamma)}{8u}\right.\\
&+\left.\frac{3-\cos\gamma}{8u^3}\right]\csc^3\left(\frac{\gamma}{2}\right)
+\left[\left(\frac{4}{u}-\frac{8}{u^3}\right)\sin^2\left(\frac{\gamma}{2}\right)+\text{Si}\left[2u+2u\sin\left(\frac{\gamma}{2}\right)\right]
\right.\\
&\left.
-2\text{Si}(2u)+\text{Si}\left[2u-2u\sin\left(\frac{\gamma}{2}\right)\right]\right]\sin(2u)+\left[\left(\frac{8}{u^2}
-\frac{8}{3}\right)\sin^2\left(\frac{\gamma}{2}\right)-2\text{Ci}(2u)
\right.\\
&\left.\left.+\text{Ci}\left[2u+2u\sin\left(\frac{\gamma}{2}\right)\right]+\text{Ci}\left[2u-2u\sin\left(\frac{\gamma}{2}\right)\right]
-\ln\left[\cos^2\left(\frac{\gamma}{2}\right)\right]\right]\cos(2u)\right\}\\
&-\frac{1}{2u^2}\left[-5+2\ln 2+2\gamma_E+2\ln (u)-\frac{1}{3}\cos(2u)-2\text{Ci}(2u)\right.\\
&\left.+\frac{-4\sin(2u)+4[1+\cos^2(u)]u+2\sin(2u)u^2}{u^3}\right],
\end{split}
\end{equation}
where $\gamma_E$ is the Euler number.

\begin{equation}
\label{mib}
\begin{split}
B(u)=&\frac{3-\cos\gamma}{12u^2}+\frac{-1+\cos\gamma}{u^4}
+\sin(2u)\left(\frac{2}{u^5}-\frac{1}{u^3}\right)\sin ^2\left(\frac{\gamma }{2}\right)\\
&+\sin\left[2u\sin\left(\frac{\gamma }{2}\right)\right]\csc ^3\left(\frac{\gamma }{2}\right) \left[\frac{\cos\gamma -3}{32u^5}+\frac{3-4 \cos\gamma+\cos (2 \gamma )}{32u^3}\right]\\
&+\cos(2u)\left(\frac{1}{6u^2}-\frac{2}{u^4}\right)\sin^2\left(\frac{\gamma}{2}\right)
+\cos \left[2 u \sin \left(\frac{\gamma }{2}\right)\right]
\frac{1+\csc ^2\left(\frac{\gamma }{2}\right)}{8 u^4}\\
&-\frac{1}{2u^2}\left[[1+\cos^2(u)](\frac{1}{3}-\frac{2}{u^2}
)+\frac{2}{u^3}\sin(2u)+\sin^2(u)\right].
\end{split}
\end{equation}

\begin{equation}
\label{rl1}
\begin{split}
A_l(u)=&\frac{v_{gw}^4}{u^3}\sin(u)\cos\left(\frac{u}{v_{gw}}\right)
-\frac{v_{gw}^5}{u^4}\sin(u)\sin\left(\frac{u}{v_{gw}}\right)
-\frac{2v_{gw}^4}{u^4}\cos(u)\cos\left(\frac{u}{v_{gw}}\right)\\
&+\frac{v_{gw}^2}{12u^2}(3v_{gw}^2+1)\cos(2u)-\frac{v_{gw}^3}{u^5}(2u^2v_{gw}^2-2v_{gw}^2+u^2)\cos(u)\sin\left(\frac{u}{v_{gw}}\right)\\
&+\frac{v_{gw}^2}{4(1-v_{gw}^2)u^2}\left[1-9v_{gw}^4+6v_{gw}^2\right.\\
&\qquad\qquad\qquad \left.+2v_{gw}^5\sin(u)\sin(\frac{u}{v_{gw}})+2v_{gw}^4\cos(u)\cos(\frac{u}{v_{gw}})\right]\\
&+\frac{\cos (2 u)+9}{8 u^2}v_{gw}^5 \left[\text{Ci} \left(u+\frac{u}{v_{gw}}\right)-\text{Ci} \left(\frac{u}{v_{gw}}-u\right)+\ln \frac{1-v_{gw}}{1+v_{gw}}\right]\\
&+\frac{ u+\sin (u)\cos(u)}{4u^2}v_{gw}^5\left[\text{Si}\left(\frac{u}{v_{gw}}+u\right)
-\text{Si}\left(u-\frac{u}{v_{gw}}\right)\right].
\end{split}
\end{equation}

\begin{equation}
\label{mil}
\begin{split}
L(u)=&\left\{\frac{13}{8}-\frac{7\cos\gamma}{12}+\frac{-1+\cos\gamma }{u^2}+\frac{u}{4}\text{Si}(2u)+\left[-\frac78+\frac{\cos\gamma}{4}+\frac{\csc^2\left(\frac{\gamma}{2}\right)}{8}\right]\left[\gamma_E\right.\right.\\
&\left.-\text{Ci}(2u)+\ln(2u)\right]
-\frac14\left[\gamma_E-\text{Ci}\left[2u\sin\left(\frac{\gamma}{2}\right)\right]+\ln\left[2u\sin\left(\frac{\gamma}{2}\right)\right]\right]\csc^2\left(\frac{\gamma}{2}\right)\\
&+\frac{1+\csc^2\left(\frac{\gamma}{2}\right)}{8u^2}\cos\left[2u\sin\left(\frac{\gamma}{2}\right)\right]+\sin\left[2u\sin\left(\frac{\gamma}{2}\right)\right]\csc^3\left(\frac{\gamma}{2}\right)\left[\frac{-3+\cos\gamma}{32u^3}\right.\\
&\left.+\frac{-5+4\cos\gamma+\cos2\gamma}{32u}\right]+\frac12\cos\gamma\cot^2\gamma\left[\sin\left[2u\sin^2\left(\frac{\gamma}{2}\right)\right]\left(\text{Si}\left[2u\sin^2\left(\frac{\gamma}{2}\right)\right]\right.\right.\\
&+\text{Si}\left[2u\sin\left(\frac{\gamma}{2}\right)-2u\sin^2\left(\frac{\gamma}{2}\right)\right]-\text{Si}\left[2u\sin\left(\frac{\gamma}{2}\right)+2u\sin^2\left(\frac{\gamma}{2}\right)\right]\\
&\left.-\text{Si}\left[2u\cos^2\left(\frac{\gamma}{2}\right)\right]\right)+\cos\left[2u\sin^2\left(\frac{\gamma}{2}\right)\right]\Big(\text{Ci}\left[2u\sin^2\left(\frac{\gamma}{2}\right)\right]+\text{Ci}\left[2u\cos^2\left(\frac{\gamma}{2}\right)\right]\\
&\left.-\text{Ci}\left[2u\sin\left(\frac{\gamma}{2}\right)-2u\sin^2\left(\frac{\gamma}{2}\right)\right]-\text{Ci}\left[2u\sin\left(\frac{\gamma}{2}\right)+2u\sin^2\left(\frac{\gamma}{2}\right)\right]\Big)\right]\\
&+\frac{\sec^2\left(\frac{\gamma}{2}\right)}{16}\left[\left(\frac{8}{u^3}-\frac{8}{u}\right)\sin^2\gamma-2\text{Si}\left[2u-2u\sin\left(\frac{\gamma}{2}\right)\right]-2\text{Si}\left[2u+2u\sin\left(\frac{\gamma}{2}\right)\right]\right.\\
&+[4+\cos\gamma-\cos(2\gamma)]\text{Si}(2u)\Big]\sin(2u)+\frac{\sec^2\left(\frac{\gamma}{2}\right)}{16}\left[\frac{10+3\cos\gamma-7\cos(2\gamma)}{3}\right.\\
&+\frac{-4+4\cos2\gamma}{u^2}-[\cos\gamma-\cos(2\gamma)]\gamma_E+[4+\cos\gamma-\cos(2\gamma)]\left[\text{Ci}(2u)-\ln(2u)\right]\\
&-\left.\left.2\text{Ci}\left[2u-2u\sin\left(\frac{\gamma}{2}\right)\right]-2\text{Ci}\left[2u+2u\sin\left(\frac{\gamma}{2}\right)\right]+4\ln\left[2u\cos\left(\frac{\gamma}{2}\right)\right]\right]\cos(2u)\right\}/u^2\\
&-\frac{1}{8u^2}\left[15-9\ln 2-9\gamma_E-9\ln(u)+\left(\frac{11}{3}-\ln 2-\gamma_E-\ln (u)\right)\cos(2u)\right.\\
&\left.+[9+\cos(2u)]\text{Ci}(2u)+[2u+\sin(2u)]\text{Si}(2u)+\frac{8[\sin(2u)-(1+\cos^2(u))u-\sin(2u)u^2]}{u^3}\right].
\end{split}
\end{equation}

\end{document}